\documentclass[fleqn,10pt]{wlscirep}
\usepackage[utf8]{inputenc}
\usepackage[T1]{fontenc}
\newcommand{\be}{\begin{equation}}
\newcommand{\ee}{\end{equation}}
\newcommand{\ra}{\rangle}
\newcommand{\la}{\langle}
\newcommand{\bea}{\begin{eqnarray}}
\newcommand{\eea}{\end{eqnarray}}
\newcommand{\nn}{\nonumber}
\title{Quantum dynamics of a driven parametric oscillator in a Kerr medium}

\author[1]{E. Bolandhemmat}
\author[1,*]{F. Kheirandish}
\affil[1]{Department of Physics, University of Kurdistan, P.O.Box 66177-15175, Sanandaj, Iran}

\affil[*]{f.kheirandish@uok.ac.ir}

\begin{abstract}
In this paper, we first analyze a parametric oscillator with both mass and frequency time-dependent. We show that the evolution operator can be obtained from the evolution operator of another parametric oscillator with a constant mass and time-dependent frequency followed by a time transformation $t\rightarrow\int_0^t dt'\,1/m(t')$. Then we proceed by investigating the quantum dynamics of a parametric oscillator with unit mass and time-dependent frequency in a Kerr medium under the influence of a time-dependent force along the motion of the oscillator. The quantum dynamics of the time-dependent oscillator is analyzed from both analytical and numerical points of view in two main regimes: (i) small Kerr parameter $\chi$, and (ii) small confinement parameter $k$. In the following, to investigate the characteristics and statistical properties of the generated states, we calculate the autocorrelation function, the Mandel $Q$ parameter, and the Husimi $Q$-function.
\end{abstract}
\begin{document}

\flushbottom
\maketitle
%
%
\thispagestyle{empty}

%
\section*{Introduction}
\noindent A momentous concept of coherent states with the eigenvalue relation $\hat{a}|\alpha\rangle = \alpha|\alpha\rangle$ as, a very convenient foundation for studying and describing the radiation field, was first introduced by Schr\"{o}dinger in $1926$ which appeared from the investigation of the quantum harmonic oscillator \cite{schrodinger1926stetige,ref1,ref2,de1996nonlinear}. But the quantum theory of coherence based on coherent states and photodetection had been developed by Glauber, Wolf, Sudarshan, Mandel, Klauder, and many others in the early $1960s$ that are most resembling quantum states in classical radiation fields and are therefore considered as the boundary between classical mechanics and quantum mechanics. Glauber innovative work was acknowledged by awarding him the Nobel Prize in $2005$ \cite{ref3,klauder1960action}. Indeed,  coherent states have become one of the most commonly used instruments in quantum physics which performed a very significant role in various fields particularly in quantum optics and quantum information. The coherent states allowed us to describe the behavior of light in phase space, using the quasi-probabilities developed much earlier by Wigner and others \cite{smith2007optics}. The significance of coherent states is because of their generalizations that have been demonstrated to have the capacity to present non-classical radiation field characteristics \cite{ghosh1998generalized,iqbal2011generalized,zhang1990coherent}.
The manifestation of the laser as a great potential coherent light marked the start of an extensive study of non-linear interactions between light and matter \cite{roman2015parametric}. This can be attained experimentally by crossing a coherent state via a Kerr medium as a result of the advent of recognizable macroscopic superpositions of coherent states, the so-called cat states \cite{de1996even}. Kerr states as the output of a Kerr medium had been introduced by Kitagawa and Yamamoto, when the state at the way in of the Kerr medium is a canonical coherent state \cite{kitagawa1986number}. The Kerr effect generates quadrature squeezing but does not modify the input field photon statistics, i.e. it remains Poissonian, which is a characteristic of the canonical coherent state input and was used for the generation of a superposition of coherent states \cite{honarasa2012generalized, tanas1991squeezing, stobinska2007effective}. Here it is worth noting that diffusion of light in a Kerr medium is also characterized by the anharmonic oscillator sample and the anharmonic term is taken to be equal to $\hat{n}^p$, where $p$ is an integer ($p> 1$) \cite{yurke1986generating, milburn1986quantum}. This oscillator mode can be evaluated as describing the evolution of a coherent state injected into a transmission line with a nonlinear susceptibility, an optical fiber for example. A laser beam that is quantum mechanically depicted by a coherent state, while passing via non-linear media, can undergo a diversity of complex alterations containing collapses and revivals of the quantum state. In any evolution of linear or non-linear, dissipation is always ready. The dissipative effects classically conduce to decreasing in the amplitude, however, if the interactions befall at atomic scales, quantum effects are significant \cite{kaur2018effect}. Nonlinear coherent states are one of the most prominent generalizations of the standard coherence states \cite{sivakumar2000studies}.

An appropriate question has been appointed: What will occur if the temporal evolution of an initial coherent state is influenced by a time-dependent harmonic-oscillator Hamiltonian with the coupling of time-dependent external additive potentials \cite{menouar2010alternative,han1999illustrative,puri1994m,lewis1967classical}? There are miscellaneous sorts of time-dependent harmonic oscillators such
as parametric oscillators \cite{ng1997coherent,roman2015parametric}, Caldirola–Kanai oscillators \cite{kanai1948quantization,caldirola1941quantum}, and harmonic oscillators with a strongly pulsating mass \cite{qian1991exact}.

Here we first investigate the quantum dynamics of a parametric oscillator with both mass and frequency time-dependent Eq. (\ref{MF1}) and show that the corresponding time-evolution operator can be obtained from another parametric oscillator with a constant mass and time-dependent frequency followed by a time transformation $t\rightarrow\int_0^t dt'\,1/m(t')$. Therefore, we mainly focus on a parametric oscillator described by the Hamiltonian $\hat{H}(t)=\hat{p}^2/2+\omega^2(t)\hat{q}^2/2$, in a Kerr medium and under the influence of a classical external source.
%
\section*{Quantum harmonic oscillator with both mass and frequency time-dependent}
\noindent To set the stage, let us first consider a parametric oscillator with time-dependent mass and frequency
\be\label{MF1}
  \hat{H}(t)=\frac{\hat{p}^2}{2m(t)} + \frac{1}{2}\,m(t)\Omega^2(t)\hat{q}^2.
\ee
The Hamiltonian Eq. (\ref{MF1}) can be written as
\be\label{MF2}
  \hat{H}(t)=\frac{1}{m(t)}\bigg(\underbrace{\frac{\hat{p}^2}{2} + \frac{1}{2}\,\omega^2(t)\hat{q}^2}_{\hat{H}^{*}(t)}\bigg),
\ee
where we have defined $\omega(t)=m(t)\Omega(t)$. Let $E_n^{*}(t)$ and $\psi_n^{*}(q,t)$ be eigenvalues and eigenfunctions of the Hamiltonian $\hat{H}^{*}(t)$ respectively
\be\label{MF3}
  \hat{H}^{*}(t)\,\psi_n^{*}(q,t)=E_n^{*}(t)\,\psi_n^{*}(q,t),
\ee
then, one easily finds
\bea\label{MF4}
&&  E_n^{*}(t)=\hbar\omega(t)\,(n+1/2),\nn\\
&&  \psi_n^{*}(q,t)=\frac{1}{\sqrt{2^n n!}}\,(\omega(t)/\pi)^{1/4}\,e^{-\frac{1}{2\hbar}\omega(t)q^2}\,H_n (\sqrt{\omega(t)/\hbar}\,q).
\eea
Therefore, from Eq. (\ref{MF2}) we deduce that the Hamiltonian $\hat{H}(t)$ has the same eigenfunctions as $\hat{H}^{*}(t)$ but the corresponding eigenvalues $E_n (t)$ are given by
\be\label{MF5}
  E_n (t)=\frac{1}{m(t)}\,E_N^{*}(t)=\hbar\Omega(t)\,(n+1/2),
\ee
that is time-dependent mass does not affect the eigenvalues but eigenfunctions, as expected. Now let us find a connection between the time evolution operators of the Hamiltonians $\hat{H}(t)$ and $\hat{H}^{*}(t)$.

Let $\hat{U} (t)$ be the time-evolution operator corresponding to $\hat{H}(t)=\hat{H}^* (t)/m(t)$, then by using
\be\label{MF6}
  \hat{U}(t)=\mbox{T}\left[e^{-\frac{i}{\hbar}\int_0^t dt'\,\frac{1}{m(t')}\hat{H}^* (t')}\right],
\ee
we deduce that if we define a new variable $\tau$ as $\tau=\varrho(t)=\int_0^t 1/m(t')\,dt'$ which is an increasing function of $t$ for $m(t)>0$, then the time-ordering will does not change and we can rewrite Eq. (\ref{MF6}) as
\be\label{MF7}
  \hat{U}(t)=\mbox{T}\left[e^{-\frac{i}{\hbar}\int_0^{\tau (t)} d\tau\,\hat{H}^* (t(\tau))}\right].
\ee
Therefore, as the first step, we replace the parameter $t$ in the Hamiltonian $\hat{H}^* (t)$  with $\varrho^{-1}(\tau)$, and define the transformed Hamiltonian as $\tilde{\hat{H}}^* (\tau)=\hat{H}^* (\varrho^{-1}(\tau))$. Let us denote the corresponding time-evolution operator by $\tilde{\hat{U}}^* (\tau)$, then
\bea\label{MF8}
i\hbar\,\frac{d}{d\tau}\tilde{\hat{U}}^*(\tau)=\tilde{\hat{H}}^* (\tau)\,\tilde{\hat{U}}^*(\tau).
\eea
Now we can prove that the time-evolution operator $\hat{U}(t)$ corresponding to the Hamiltonian $\hat{H}(t)$ can be obtained from $\tilde{\hat{U}}^* (\tau)$ by replacing $\tau$ with $\varrho(t)=\int_0^t 1/m(t')\,dt'$, that is $\hat{U}(t)=\tilde{\hat{U}}^* (\varrho(t))$. We have
\bea\label{prove}
i\hbar\,\frac{d}{dt}\hat{U}(t)=&& i\hbar\,\frac{d}{dt}\tilde{\hat{U}}^* (\varrho(t))=i\hbar\,\frac{d\varrho}{dt}\frac{d}{d\varrho}\tilde{\hat{U}}^* (\varrho)=\frac{1}{m(t)}i\hbar\,\frac{d}{d\varrho}\tilde{\hat{U}}^* (\varrho),\nn\\
                              =&& \frac{1}{m(t)}\,\tilde{\hat{H}}^{*}(\varrho)\,\tilde{\hat{U}}^* (\varrho)=\frac{1}{m(t)}\,\hat{H}^{*}(t)\,\tilde{\hat{U}}^* (\varrho)=\hat{H}(t)\,\hat{U}(t).
\eea
%
\subsection*{Example}
\noindent As an example let us find the time-evolution operator for the Hamiltonian
\bea\label{ex1}
\hat{H}(t)=&& \frac{\hat{p}^2}{2m_0\exp(\gamma t)}+\frac{1}{2}m_0\exp(\gamma t)\,\omega_0^2 \,\hat{q}^2,\nn\\
          =&& \frac{1}{\exp(\gamma t)}\left(\frac{\hat{p}^2}{2m_0}+\frac{1}{2}m_0\omega_0^2 \exp(2\gamma t)\,\hat{q}^2\right),
\eea
we have
\bea
&& \tau=\int_0^t dt'\,\frac{1}{\exp(\gamma t')}=\frac{1}{\gamma}(1-\exp(-\gamma t)),\nn\\
&& \tilde{\hat{H}}^* (\tau)=\frac{\hat{p}^2}{2m_0}+\frac{1}{2}m_0\omega_0^2 \frac{1}{(1-\gamma \tau)^2}\,\hat{q}^2.\label{pro}
\eea
The quantum propagator for Hamiltonians of type Eq. (\ref{pro}) has been investigated in \cite{kheirandish2018novel}, let us denote the quantum propagator of $\tilde{\hat{H}}^* (\tau)$ in position space by $\tilde{K}^*(q,\tau|q',0)$, then the quantum propagator corresponding to the main Hamiltonian $\hat{H}(t)$ is
\be\label{proposition}
K(q,t|q',0)= \tilde{K}^*(q,\frac{1}{\gamma}(1-e^{-\gamma t})|q',0).
\ee
Also, the position and momentum operators in the Heisenberg picture are given by
\bea
  \hat{q}(t) =&& \hat{U}^\dag (t)\,\hat{q}(0)\,\hat{U}(t)={\tilde{\hat{U}}^*}^\dag (\tau)\,\hat{q}(0)\, {\tilde{\hat{U}}^*(\tau)}=\hat{q}^*(\tau)\big|_{\tau=\varrho(t)},\nn\\
  \hat{p}(t) =&& \hat{U}^\dag (t)\,\hat{p}(0)\,\hat{U}(t)={\tilde{\hat{U}}^*}^\dag (\tau)\,\hat{p}(0)\, {\tilde{\hat{U}}^*(\tau)}=\hat{p}^*(\tau)\big|_{\tau=\varrho(t)},
\eea
where $\hat{q}^*(\tau)$ and $\hat{p}^*(\tau)$ are position and momentum operators in the Heisenberg picture corresponding to the Hamiltonian $\tilde{\hat{H}}^*(\tau)$. From the Heisenberg equation for $\hat{q}^*(\tau)$ one finds
\bea\label{qstar}
&& \frac{d^2}{d\tau^2}\,\hat{q}^*(\tau)+\frac{\omega_0^2}{(1-\gamma \tau)^2}\,\hat{q}^*(\tau)=0,\nn\\
&& \hat{p}^*(\tau)=m_0\,\frac{d}{d\tau}\,\hat{q}^*(\tau).
\eea
The Eqs. (\ref{qstar}), have the following solutions
\bea
\hat{q}^*(\tau)=&& \hat{C}_1\,(\tau-1/\gamma)^{\frac{\gamma+i\digamma}{2\gamma}}+\hat{C}_2\,(\tau-1/\gamma)^{\frac{\gamma-i\digamma}{2\gamma}},\nn\\
\hat{p}^*(\tau)=&& \hat{C}_1\,(\frac{\gamma+i\digamma}{2\gamma})(\tau-1/\gamma)^{-\frac{\gamma-i\digamma}{2\gamma}}+\hat{C}_2\,\frac{\gamma-i\digamma}{2\gamma}(\tau-1/\gamma)^{-\frac{\gamma+i\digamma}{2\gamma}},
\eea
where $\digamma=\sqrt{4\omega_0^2-\gamma^2}$, $\tan\theta=\digamma/\gamma$, and the constant operators $\hat{C}_1$ and $\hat{C}_2$ can be obtained from the initial conditions $\hat{q}^*(0)=\hat{q}(0)$ and $\hat{p}^*(0)=\hat{p}(0)$.
After straightforward calculations, we obtain
\bea
&& \hat{q}(t)=\hat{q}^*(\tau)\big|_{\tau=\varrho(t)}=\frac{2}{m_0 \digamma}\,e^{-\frac{\gamma t}{2}}\big(\sin(\digamma t/2)\,\hat{p}(0)+m_0\omega_0\,\sin(\digamma t/2+\theta)\,\hat{q}(0)\big),\nn\\
&& \hat{p}(t)=\hat{p}^*(\tau)\big|_{\tau=\varrho(t)}=-\frac{2\omega_0}{\digamma}\,e^{-\frac{\gamma t}{2}}\big(\sin(\digamma t/2-\theta)\,\hat{p}(0)+m_0\omega_0\,\sin(\digamma t/2)\,\hat{q}(0)\big).
\eea
Therefore, the Hamiltonian $\hat{H}^*(t)$ is the main ingredient in Eq. (\ref{MF2}). In the next section, we will focus on the Hamiltonians of the type $\hat{H}^*(t)$ in the presence of an external time-dependent classical source in a Kerr medium.
\section*{The model}
\noindent The model that we will investigate in the following is a generalization of the Hamiltonian Eq. (\ref{MF2}) given by
\be\label{H}
\hat{H}(t)=\frac{1}{2}[\hat{p}^2 + {{\Omega}^2(t)\hat{q}^2}]+e(t)\hat {q}+\hat {H}_{kerr},
\ee
describing the quantum dynamics of a time-dependent harmonic oscillator in a Kerr medium and under the influence of a time-dependent force $-e(t)$ along the motion of the oscillator. In Eq. (\ref{H}), $\Omega(t)$ is a time-dependent frequency and $\hat {H}_{kerr} ={\chi\hat {n}^2}$. The Kerr parameter $\chi$ is a constant proportional to the third-order nonlinear susceptibility $\chi^{3}$ which is, in general, a small parameter. To be specific, in what follows we will choose $\Omega(t)=\Omega_0[1+2k\cos(2\Omega_0 t)]$ where $k$ is also a small confinement parameter \cite{de2015generation}. To this end, the annihilation, creation, and number operators are defined respectively by
\bea\label{a}
\hat{A_t}&=&\frac{1}{\sqrt{2\Omega(t)}}(\Omega(t)\hat{q}+i\hat{p}),\nonumber\\
\hat {A}^\dagger_t &=& \frac{1}{\sqrt{2\Omega(t)}}(\Omega(t)\hat{q}-i\hat{p}),\nonumber\\
\hat{n}_t &=& \hat {A}^\dagger_t \hat{A_t},
\eea
where for notational simplicity we have set $\hbar=1$. The time-dependent operators given in Eq. (\ref{a}) fulfill the Heisenberg algebra at any time
\bea\label{Halg}
&& [\hat{A_t},\hat {A}^\dagger_t] = 1,\nonumber\\
&& [\hat{n}_t,\hat{A}^\dagger_t] = \hat{A}^\dagger_t,\nonumber\\
&& [\hat{n}_t,\hat{A_t}] = \hat{A_t}.
\eea
In the absence of a Kerr medium ($\chi=0$), the Hamiltonian Eq. (\ref{H}) reduces to $\hat{H}_f(t)$ given by
\bea\label{H0}
\hat{H}_f(t) &=& \frac{1}{2}[\hat{p}^2 + {{\Omega}^2(t)\hat{q}^2}]+e(t)\hat{q},\nn\\
&=& \underbrace{\Omega(t)(\hat {A}^\dagger_t \hat{A_t}+ \frac{1}{2})}_{\hat{H}_0 (t)}+\frac{e(t)}{\sqrt{2\Omega(t)}}(\hat{A}^\dag_t+\hat{A}_t).
\eea
The Hamiltonian $\hat{H}_f(t)$ can be diagonalized. To this end, let us define the time-dependent displacement operator as
\be\label{Disop}
\hat{D}_t (\alpha)=e^{\alpha \hat{A}^\dag_t-\bar{\alpha}\hat{A}_t}.
\ee
By making use of the Baker-Campbell-Hausdorff (BCH) formula we find
\bea\label{DD}
&&\hat{D}^\dagger_t(\alpha)(\hat{A}^\dagger_t)\hat{D}_t(\alpha)=\hat{A}^\dagger_t+\bar{\alpha},\nn\\
&&\hat{D}^\dagger_t(\alpha)(\hat{A}_t)\hat{D}_t(\alpha)=\hat{A}_t+\alpha,
\eea
also
\bea\label{H1}
\hat{D}^\dagger_t(\alpha)\hat{H}_0(t)\hat{D}_t(\alpha)&=& \Omega_t\Big[(\hat {A}^\dagger_t\hat{A_t}+\frac{1}{2})+\alpha\hat {A}^\dagger_t+\bar{\alpha}\hat{A_t}+|\alpha|^2\Big],\nn\\
&=& \hat{H}_f(t)+\frac{e^2 (t)}{2\Omega^2 (t)},
\eea
where $\hat{H}_0(t)$ is given in Eq. (\ref{H0}) and for convenience we defined
\be
\alpha=\bar{\alpha}=\lambda_t=\frac{e(t)}{\Omega(t)\sqrt{2\Omega(t)}}.
\ee
Therefore, the Hamiltonian $\hat{H}_f(t)$ is obtained from $\hat{H}_0 (t)$ trough a similarity transformation followed by a translation as
\bea\label{main}
\hat{H}_f (t) &=& \hat {D}^\dagger_t(\lambda_t)\hat{H}_0 (t)\hat {D_t}(\lambda_t)-\Omega(t)\,\lambda_t^2.
\eea
Let $|n\ra^0_t$ and $E^0_n (t)$ be the eigenstates and eigenvalues of the Hamiltonian $\hat{H}_0 (t)$ respectively
\bea\label{h0eigen}
\hat{H}_0 (t)|n\ra^0_t &=& E^0_n (t) |n\ra^0_t,\,\,n=0,1,2,\cdots,\nn\\
E^0_n (t) &=& (n+1/2)\Omega(t),
\eea
by using Eq. (\ref{main}) one easily finds that the states $|n\ra_t=D^{\dag}_t(\lambda_t)|n\ra^0_t$ are the eigenstates of the Hamiltonian $\hat{H}_f (t)$ with eigenvalues $E_n (t)$
\bea\label{x0}
\hat{H}_f (t)|n\ra_t &=& E_n (t)|n\ra_t,\nn\\
E_n (t) &=& E^0_n (t)-\Omega(t)\,\lambda_t^2,\nn\\
        &=& (n+1/2-\lambda_t^2)\,\Omega(t).
\eea
%
\subsection*{Position representation of the eigenfunctions of $\hat{H}_f (t)$}
\noindent The position representation of the eigenfunctions of the Hamiltonian $\hat{H}_f (t)$ can be obtained as follows
\bea\label{e1}
\psi_n^f (q,t)=\la q|n\ra_t=&& \la q|\hat{D}^\dag_t (\lambda_t)|n\ra^0_t,\nn\\
                           =&& \la q|e^{\lambda_t(\hat{A}_t-\hat{A}^\dag_t)}|n\ra^0_t,\nn\\
                           =&& \la q|e^{i\lambda_t\sqrt{\frac{2}{\Omega(t)}}\hat{p}}|n\ra^0_t,\nn\\
                           =&& \la q-\lambda_t\sqrt{2/\Omega(t)}||n\ra^0_t,\nn\\
                           =&& \psi_n^0 (q-\lambda_t\sqrt{2/\Omega(t)},t),
\eea
where we made use of Eqs. (\ref{a}). The eigenfunction $\psi_n^0 (q,t)$ of the Hamiltonian $\hat{H}_0 (t)$ can be obtained from $\psi_n^0 (q,t)=(\hat{A}^\dag_t)^n/\sqrt{n!}\,\psi_0^0 (q,t)$, where $\hat{A}_t\,\psi_0^0 (q,t)=0$, the explicit form of the eigenfunction $\psi_n^0 (q,t)$ is
\begin{equation}\label{sit0}
  \psi_n^0 (q,t)=\la q|n\ra_t^0=\frac{1}{\sqrt{2^n n!}}\left(\frac{\Omega(t)}{\pi}\right)^{1/4}\,e^{-\frac{1}{2}\Omega(t)q^2}\,H_n (\sqrt{\Omega(t)}\,q),
\end{equation}
where $H_n (z)$ is a Hermite polynomial of order $n$
\begin{equation}\label{hermite}
  H_n (z)=(-1)^n e^{z^2}\frac{d^n}{dz^n}e^{-z^2}.
\end{equation}
Therefore, in the presence of an external source $(\lambda_t\neq 0)$, the eigenfunction $\psi_n^f (q,t)$ is obtained by shifting $q\rightarrow q-\lambda_t\sqrt{2/\Omega(t)}$ in the free eigenfunction $\psi_n^0 (q,t)$.
\subsection*{Linearization of the Hamiltonian}
\noindent In this section, in the framework of the Heisenberg picture, we will find approximate solutions for the time-evolution of the ladder operators $\hat{a} (t)$ and $\hat{a}^\dag (t)$ using a linearization process. For this purpose, we assume that the confinement parameter is negligible ($k\ll 1$), so $\Omega(t)\approx \Omega_0$. The time-dependent Hamiltonian $\hat{H}(t)$ now becomes
\bea\label{hft}
\hat{H}_{k=0} (t)=\Omega_0\,(\hat{a}^\dag\hat{a}+1/2)+\frac{e(t)}{\sqrt{2\Omega_0}}\,(\hat{a}^\dag+\hat{a})+\chi\,(\hat{a}^\dag\hat{a})^2,
\eea
where
\bea
&& \hat{a}=\hat{A}_0=\frac{1}{\sqrt{2\Omega_0}}\,(\Omega_0\,\hat{q}+i\hat{p}),\nn\\
&& \hat{a}^\dag=\hat{A}^\dag_0=\frac{1}{\sqrt{2\Omega_0}}\,(\Omega_0\,\hat{q}-i\hat{p}),\nn\\
&& \hat{n}=\hat{A}^\dag_0\hat{A}_0=\hat{a}^\dag\hat{a}.
\eea
From Heisenberg equation we have
\bea\label{HK}
i\dot{\hat{a}}&=&[\hat{a},\hat{H}_{k=0}(t)],\nonumber\\
              &=&\nu\,\hat{a}+2\chi\hat{a}\hat{n}+\frac{e(t)}{\sqrt{2\Omega_0}},
\eea
where $\nu=\Omega_0-\chi$. By inserting $\hat{a}(t)=e^{-i\nu t}\,\hat{b}(t)$ into Eq. (\ref{HK}) we find
\bea\label{HKb}
&&\dot{\hat{b}}(t)=-2i\chi\,\hat{b}(t)\,\hat{n}_t-i\frac{e(t)}{\sqrt{2\Omega_0}}e^{-i\nu t}.
\eea
In Eq. (\ref{HK}) the term $2\chi \hat{a}\hat{n}$ can be ignored up to the first order approximation since $\chi\ll 1$, then
\bea\label{k0aadog}
\hat{a}(t) & \approx& e^{-i\nu t}\hat{a}(0)-\zeta(t),\nonumber\\
\hat{a}^\dag(t) & \approx& e^{i\nu t}\hat{a}^\dag(0)-\bar{\zeta}(t),
\eea
where
\be
\zeta(t)=i\frac{1}{\sqrt{2\Omega_0}}\int_{0}^{t}dt^\prime e^{-i\nu (t-t^\prime)}e(t^\prime).
\ee
To proceed, let the initial state of the system be a number state $|\psi(0)\rangle=|n\rangle$ then we can linearize Eq. (\ref{HKb}) by replacing $\hat{n}_t$ with its average value $\bar{n}(t)$
\bea\label{nn}
\bar{n}(t)&&=\langle n|\hat{a}^\dag(t)\hat{a}(t)|n\rangle,\nonumber\\
&&=\langle n|\hat{b}^\dag(t)\hat{b}(t)|n\rangle,\nonumber\\
&&=\langle n|\hat{a}^\dag(0)\hat{a}(0)-e^{i\nu t}\varphi_t\hat{a}^\dag(0)-e^{-i\nu t}\varphi^\star(t)\hat{a}(0)+|\varphi_t|^2|n\rangle,\nonumber\\
&&=n+|\varphi_t|^2,
\eea
we find
\bea\label{eq}
\dot{\hat{b}}(t)=-2i\chi\hat{b}(t)[n+|\varphi_t|^2]-i\frac{e(t)}{\sqrt{2\Omega_0}}\,e^{-i\nu t}.
\eea
The solution of Eq.(\ref{eq}) is
\bea\label{bn}
\hat{b}(t)=e^{-i\gamma(t)}\hat{b}(0)+\delta(t),
\eea
where
\bea
&& \gamma(t)=\int_{0}^{t}A(t^\prime)dt^\prime,\\
&& A(t)=2\chi[n+|\varphi_t|^2],\\
&& \delta(t)=e^{-i\gamma(t)}\int_{0}^{t}dt^\prime e^{i\gamma(t^\prime)}\frac{e(t^\prime)}{i\sqrt{2\Omega_0}}.
\eea
Now using Eq. (\ref{bn}) we find a better solution for the ladder operators
\bea
\hat{a}(t)&&=e^{-i(\nu t+\gamma(t))}\hat{a}(0)+e^{-i\nu t}\,\delta(t),\\
\hat{a}^\dag(t)&&=e^{i(\nu t+\gamma(t))}\hat{a}^\dag(0)+e^{i\nu t}\,\bar{\delta}(t).
\eea
%
\section*{Time-evolution operator}
\noindent In this section, we reconsider the Hamiltonian Eq. (\ref{hft}) and try to find the corresponding time-evolution operator approximately. For this purpose, we make use of the properties of Heisenberg algebra $\{1,\hat{a}, \hat{a}^\dag\}$. Let us rewrite the Hamiltonian Eq. (\ref{hft}) in the following form
\bea\label{hft}
\hat{H}_{k=0} (t) &=& \Omega_0\,(\hat{a}^\dag\hat{a}+1/2)+\chi\,(\hat{a}^\dag\hat{a})^2+\frac{e(t)}{\sqrt{2\Omega_0}}\,(\hat{a}^\dag+\hat{a}),\nn\\
                  &=& \hat{H}_0+\frac{e(t)}{\sqrt{2\Omega_0}}\,(\hat{a}^\dag+\hat{a}),
\eea
where $\hat{H}_0=\Omega_0 (\hat{n}+1/2)+\chi\hat{n}^2$. The evolution operator $\hat{U}(t)$ corresponding to the Hamiltonian $\hat{H}_{k=0}$ can be written as
\be
\hat{U}(t)=\hat{U}_0(t)\hat{U}_I(t),
\ee
where $\hat{U}_0(t)$ is the evolution operator corresponding to the time-independent Hamiltonian $\hat{H}_0$ given by
\bea
\hat{U}_0(t) &=& e^{-i\hat{H}_0t},\nonumber\\
             &=& e^{-i\Omega_0t(\hat{n}+\frac{1}{2})-it\chi\hat{n}^2}.
\eea
The time-dependent part $\hat{U}_I(t)$ fulfills
\be\label{Uieq}
i\frac{d\hat{U}_I(t)}{dt}=\hat{V}_I(t)\,\hat{U}_I(t),
\ee
where
\bea
\hat{V}_I(t) &=& \hat{U}_0^\dag(t)\hat{V}\hat{U}_0(t),\nn\\
             &=& \frac{e(t)}{\sqrt{2\Omega_0}}(e^{-i\Omega_0t-i\chi t(2\hat{n}+1)}\hat{a}+\hat{a}^\dag e^{i\Omega_0t+i\chi t(2\hat{n}+1)}),\nonumber\\
             &=& g(t)\hat{a}+\hat{a}^\dag \bar{g}(t),\label{gtf}
\eea
and
\be
g(t)=\frac{e(t)}{\sqrt{2\Omega_0}}\,e^{-it(\Omega_0+\chi)}e^{|\alpha|^2(e^{-2i\chi t}-1)}.
\ee
In deriving Eq. (\ref{gtf}), to simplify the calculations, we assumed that the system is initially prepared in a coherent state $|\alpha\rangle$, and replaced the nonlinear term $e^{[\pm i\Omega(\hat{n})t]}$ by its average value $\langle \alpha|e^{[\pm i\Omega(\hat{n})t]}|\alpha\rangle$, see  \cite{walls2012quantum,dodonov1998dynamical,roman2016approximate,berrondo2011dipole, gerry2005introductory}.

To find $\hat{U}_I(t)$, we assume
\bea\label{u1}
\hat{U}_I(t)=\prod_{n=1}^{3}\exp(X_n(t)\hat{\gamma}_n),
\eea
where we have defined $\hat{\gamma}_1=1, \hat{\gamma}_2=\hat{a}, \hat{\gamma}_3=\hat{a}^\dag$ and the initial conditions are $X_1(0)=X_2(0)=X_3(0)=0$.
By inserting Eq. (\ref{u1}) into Eq. (\ref{Uieq}), we find \cite{wei1964global,wei1963lie}
\bea\label{xx}
  \dot{X}_1(t)&&= \dot{X}_3(t){X}_2(t), \nn\\
  \dot{X}_2(t)&&=-i\bar{g}(t), \nn\\
  \dot{X}_3(t)&&=-i g(t).
\eea
The Eqs. (\ref{xx}) can be solved either analytically or numerically and from now on we assume that the solutions are known functions. Therefore, if we denote the temporal evolution of the initial state $|\alpha\rangle$ by $|\alpha,t\rangle$, we have
\bea\label{si}
|\psi_\alpha,t\rangle&&=\hat{U}_0(t)\hat{U}_I(t)|\alpha\rangle,\nn\\
&&=G_\alpha\sum_{n=0}^{\infty}e^{-i\Omega_0tn-itn^2\chi}\frac{(X_2(t)+\alpha)^n}{\sqrt{n!}}|n\rangle.
\eea
where $G_\alpha=e^{X_3(t)\alpha+X_1(t)-i\frac{\Omega_0t}{2}-\frac{|\alpha|^2}{2}}$, and from the normalization condition of the wave function $\langle \psi_\alpha,t |\psi_\alpha,t\rangle=1$, we find $|G_\alpha|^2=\exp(-|(X_2(t)+\alpha)|^2)$.
For convenience, let us define the dimensionless parameters $\xi=\chi t$ and $X_2(t)+\alpha=\eta_t$, then Eq. (\ref{si}) can be rewritten as
\bea\label{sif}
|\psi_\alpha,t\rangle &&=e^{\frac{-|\eta_t|^2}{2}}e^{-i\xi\hat{n}^2}\sum_{n=0}^{\infty}\frac{(e^{-i\Omega_0t}\eta_t)^n}{\sqrt{n!}}|n\rangle,\nn\\
                      &&=e^{-i\xi\hat{n}^2}|e^{-i\Omega_0t}\eta_t\rangle,\nn\\
                      &&=|e^{-i\Omega_0 t}\eta_t\rangle_\xi,
\eea
where we have defined $|\zeta\rangle_\xi=e^{-i\xi\hat{n}^2}|\zeta\rangle$ for an arbitrary coherent state $|\zeta\ra$. Therefore, the evolved state $|\psi_\alpha,t\rangle$ is of the kind $|\beta\ra_{\xi}$ where $\beta=e^{-i\Omega_0 t}\eta_t$. In the next section we will study the properties of these states.
\subsubsection*{Properties of the states $|\beta\ra_{\xi}$}
\noindent Let us define a new set of ladder operators as
\bea\label{Bs}
&& \hat{B}=\hat{a}\,f(\hat{n})=f(\hat{n}+1)\,\hat{a},\nn\\
&& \hat{B}^\dag=f^\dag(\hat{n})\,\hat{a}^\dag=\hat{a}^\dag\,f^\dag(\hat{n}+1),
\eea
where the function $f(\hat{n})$ is defined by
\be
f(\hat{n})=e^{i \xi (2 \hat{n}-1)}.
\ee
The operators $\hat{B}$, $\hat{B}^\dag$ and $\hat{n}$, fulfil the usual Heisenberg algebra
\bea
&& [\hat{B},\hat{B}^\dag]=1,\nn\\
&& [\hat{n},\hat{B}]=-\hat{B},\nn\\
&& [\hat{n},\hat{B}^\dag]=\hat{B}^\dag.
\eea
The state $|\beta\ra_{\xi}$ can be expanded in number states basis as
\bea
 |\beta\ra_{\xi} &=& e^{-i\xi\hat{n}^2}\left(e^{-\frac{|\beta|^2}{2}}\sum_{n=0}^\infty \frac{\beta^n}{\sqrt{n!}}\,|n\ra\right),\nn\\
                 &=& e^{-\frac{|\beta|^2}{2}}\sum_{n=0}^\infty \frac{\beta^n}{\sqrt{n!}}e^{-i\xi n^2}\,|n\ra,\nn\\
                 &=& e^{-\frac{|\beta|^2}{2}}\sum_{n=0}^\infty \frac{\beta^n}{\sqrt{n!}}\frac{1}{[f(n)]!}\,|n\ra,
\eea
where $[f(n)]!=\displaystyle\Pi_{k=1}^n f(n)$, and $[f(0)]!=1$. One can easily show that the state $|\beta\ra_{\xi}$ is a coherent state for the new annihilation operator $\hat{B}$ with eigenvalue $\beta$
\be
\hat{B} |\beta\ra_{\xi}=\beta\,|\beta\ra_{\xi}.
\ee
If we define the modified displacement operator $\hat{D}_B (\beta)=e^{\beta\,\hat{B}^\dag-\bar{\beta}\,\hat{B}}$, then $|\beta\ra_{\xi}=\hat{D}_B |0\ra$, note that $|0\ra_{\xi}=|0\ra$, and the parameter $\xi=\chi t$ is hidden in the definition of $\hat{B}$ and $\hat{B}^\dag$.

The state $|\beta\ra_{\xi}$ can also be considered as a Kerr state if we consider the Hamiltonian
\bea\label{KerrH}
\hat{H} &=& \hbar\chi\,\hat{a}^\dag\hat{a}+\hbar\chi\,\hat{a}^\dag\hat{a}^\dag\hat{a}\hat{a},\nn\\
        &=& \hbar\chi\hat{n}^2,
\eea
with the corresponding time-evolution operator
\bea
\hat{U} (t)=e^{-it\chi\,\hat{n}^2}=e^{-i\xi\hat{n}^2}.
\eea
If the system is initially prepared in the coherent state $|\beta\ra$, then the evolved state is the state $|\beta\ra_{\xi}$ given by
\bea
\hat{U}(t) |\beta\ra &=& e^{-i\xi\hat{n}^2}|\beta\ra=|\beta\ra_{\xi},\nn\\
                     &=& e^{-\frac{|\beta|^2}{2}}\sum_{n=0}^\infty \frac{\beta^n}{\sqrt{n!}} e^{-i\xi n^2} |n\ra.
\eea
The probability of having $n$ excitation in the evolved state $|\beta\ra_{\xi}$ is a Poissonian distribution
\bea
P_{\xi}(n)=|\la n|\beta\ra_{\xi}|^2=e^{-|\beta|^2}\frac{|\beta|^{2n}}{n!}.
\eea
To study the squeezing effects, let us find the normalized variances of the position $\hat{x}=(\hat{a}+\hat{a}^\dag)/2$ and momentum $\hat{p}= (\hat{a}-\hat{a}^\dag)/2i $ defined by $(\triangle q)_\xi/ (\triangle q)_{\xi=0}$ and $(\triangle p)_\xi/ (\triangle p)_{\xi=0}$, respectively.
We have
\begin{eqnarray}
  \frac{(\Delta q)_\xi}{(\Delta q)_{\xi=0}} &=& \sqrt{2|\beta|^2+1+2|\beta|^2\,e^{-2|\beta|^2\sin^2(2\xi)}\cos(2\phi-4\xi-|\beta|^2\sin(4\xi))-4|\beta|^2\,e^{-4|\beta|^2\sin^2(\xi)}\cos^2 (\phi-\xi-|\beta|^2\sin(2\xi))},\nn\\
 \frac{(\Delta p)_\xi}{(\Delta p)_{\xi=0}} &=& \sqrt{2|\beta|^2+1-2|\beta|^2\,e^{-2|\beta|^2\sin^2(2\xi)}\cos(-2\phi+4\xi+|\beta|^2\sin(4\xi))-4|\beta|^2\,e^{-4|\beta|^2\sin^2(\xi)}\sin^2 (\phi-|\beta|^2\sin(2\xi))}.\nn\\
\end{eqnarray}
where $\beta=|\beta|e^{i\phi}$ and $(\Delta q)_{\xi=0}=1/\sqrt{2\Omega_0}$ and $(\Delta p)_{\xi=0}=\sqrt{\Omega_0/2}$. Note that $|\beta\ra_{\xi=0}=|\beta\ra$ ia a coherent state for $\hat{a}$ leading to a minimal uncertainty. In Fig. 1, the variances are depicted for $\beta=0.5$.
\begin{figure}[ht]
\centering
\includegraphics[scale=0.4]{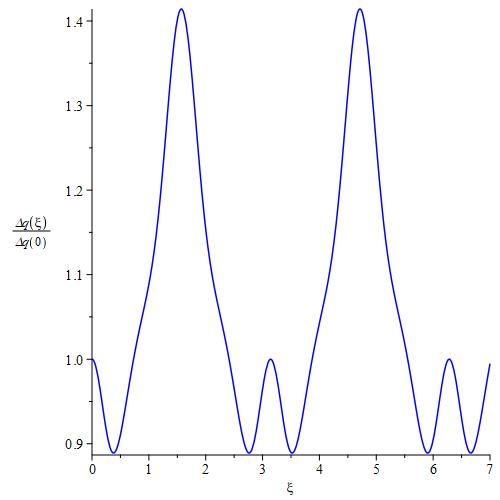}
\includegraphics[scale=0.4]{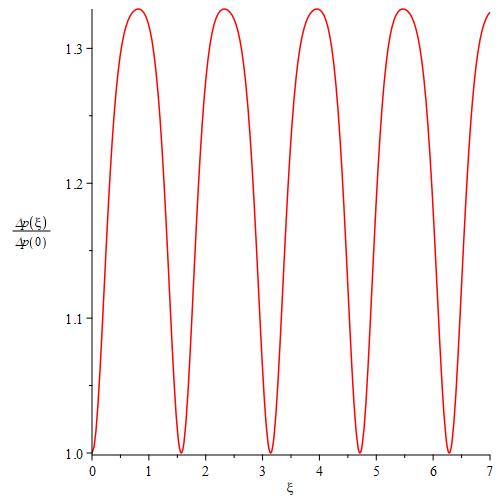}
\caption{(Color online) The normalized variances of position (left) and momentum (right) depicted for $\beta=0.5$ in dimensionless parameter $\xi=\chi t$. Squeezing occurs only for position in time intervals where $\frac{(\Delta q)_\xi}{(\Delta q)_{\xi=0}}<1$.}
\label{fig1}
\end{figure}
In the next section, we will focus on statistical properties of the state $|\beta\ra_{\xi}$ using (quasi)probability distribution functions.
\section*{Phase-space (quasi)probability distributions}
\subsection*{Mandel $Q$ parameter}
\noindent The Mandel $Q$ parameter measures the deviation of the occupation number distribution from Poissonian statistics. The quantum state $|\psi\rangle$ has a sub-Poissonian, Poissonian or super-Poissonian statistics if $Q< 0$, $Q=0$ or $Q>0$, respectively. The Mandel $Q$ parameter is defined by \cite{gerry2005introductory}
\bea
Q_M &=& \frac{\langle(\Delta\hat{n})^2\rangle-\langle\hat{n}\rangle}{\langle\hat{n}\rangle},\nonumber\\
    &=& \frac{\langle\hat{n}^2\rangle-\langle\hat{n}\rangle^2}{\langle\hat{n}\rangle}-1,\nonumber\\
    &=& \langle\hat{n}\rangle(g^{(2)}(0)-1),
\eea
where $\hat{n}$ is the photon number operator and $g^{(2)}(0)$ is the normalized second-order correlation function.

For the state $|\beta\ra_{\xi}$ we have
\bea
\la\hat{n}^2\ra_{\xi} &=& \la |\beta|e^{i\xi \hat{n}^2}\,\hat{n}^2\,e^{-i\xi \hat{n}^2}|\beta\ra=\la\beta|\hat{n}^2|\beta\ra,\nonumber\\
                      &=& |\beta|^4+|\beta|^2,\nn\\
  \la\hat{n}\ra_{\xi} &=& \la\beta|\hat{n}|\beta\ra=|\beta|^2,
\eea
therefore, $Q=0$, indicating that the statistical distribution of excitations is Poissonian. In the next section, we will study the autocorrelation function to find out how the evolved state resembles the original state.
\subsection*{Autocorrelation function}
\noindent The autocorrelation function is the overlap between the evolved and the initial state \cite{nauenberg1990autocorrelation}, and shows the possibility of total or partial resemble of the initial state when the overlap is complete or partial, respectively. The overlap or the scalar product of the initial and the evolved state $|\psi_\alpha,t\rangle$ is (see Eq. (\ref{sif})
\bea
F(t) &=& \langle\psi_\alpha,0|\psi_\alpha,t\rangle,\nonumber\\
     &=& \langle \alpha|\psi_\alpha,t\rangle,\nn\\
     &=& e^{-\frac{(|\alpha|^2+|\eta_t|^2)}{2}}\sum_{n=0}^{\infty} e^{-i\chi t n^2}\frac{(e^{-i\Omega_0t}\bar{\alpha}\eta_t)^n}{n!}.
\eea
\begin{figure}[ht]
\centering
\includegraphics[scale=0.4]{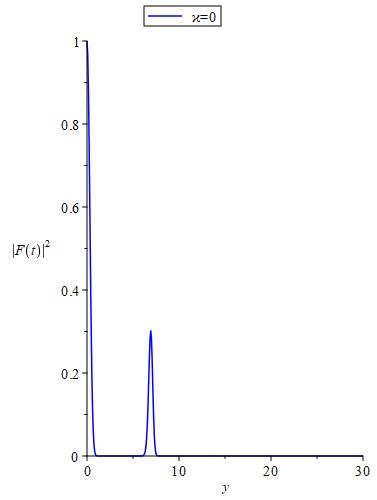}
\includegraphics[scale=0.4]{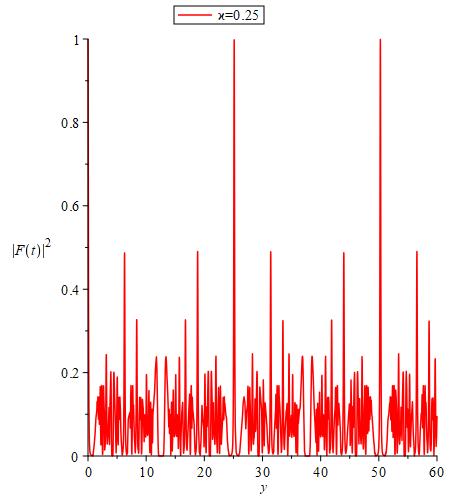}
\includegraphics[scale=0.4]{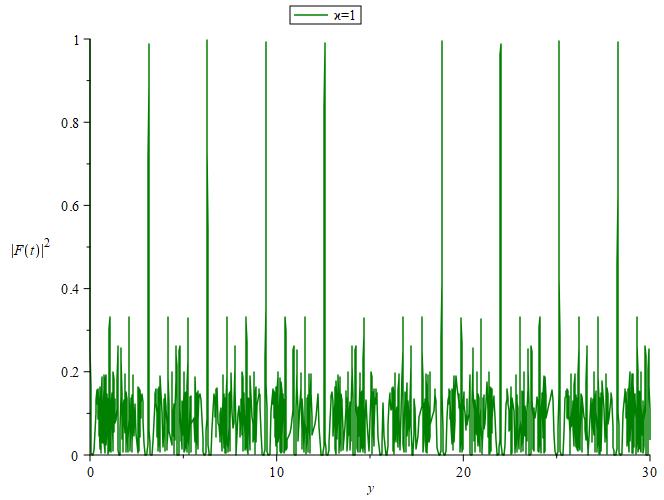}
\caption{(Color online) The function $|F(t)|^2$ in the presence of external force $e(y)=\cos(y),\,\,\,(y=\Omega_0 t)$ for the parameters: $\Omega_0=1$, $\alpha = 3$, and Kerr parameters $\chi=0$, $\chi=0.25$ and $\chi=1$.}
\label{fig1}
\end{figure}
In Fig. (\ref{fig1}), the function $|F(t)|^2$ has been depicted in the presence of the external source $e(\tau)=\cos(\tau)$ for different values of the Kerr parameter $\chi$. In the absence of the Kerr nonlinearity ($\chi=0$), we have a driven oscillator, and in this case $|F(t)|^2$ is decreasing in time without a considerable revival. In the presence of Kerr nonlinearity ($\chi=0.25$, $\chi=1$), there are periodic fractional and complete revivals with a period decreasing with increasing the Kerr parameter $\chi$.

\subsection*{Husimi distribution function}
\noindent The Husimi function, which can be measured using quantum tomographic techniques, is always positive so it is a distribution on phase space. It has been found that the Husimi distribution function is linked to classical information entropy, which can be used to measure non-classical correlations in composite systems, through the Wehrl entropy. In the phase space, the Husimi distribution function has been used to measure and study the erasing information, coherence loss, relaxation processes and adjustable phase-space information \cite{bolda1998measuring}. The Husimi function is defined by \cite{glauber1965optical}
\bea
Q(\gamma,t)=\frac{1}{\pi}\langle \gamma|\hat{\rho}(t)|\gamma\rangle.
\eea
Having the Husimi $Q$-function, we can obtain the expectation value of an arbitrary observable $A(\hat{a},\hat{a}^\dag)$ as
\bea
\la A(\hat{a},\hat{a}^\dag)\ra_t=\int d^2 \alpha\,Q(\alpha,t)\,A(\alpha,\alpha^*),
\eea
where $A(\hat{a},\hat{a}^\dag)$ is anti-normally ordered
\bea
A(\hat{a},\hat{a}^\dag)=\sum_{n,m} c_{nm}\,\hat{a}^n\hat{a}^{\dag m}.
\eea
For the pure state $\hat{\rho}(t)=|\psi_\alpha,t\rangle\langle \psi_\alpha,t|$, (see Eq. (\ref{sif})), we have
\bea
Q(\gamma,t)&&=\frac{1}{\pi}\langle \gamma|\psi_\alpha,t\rangle\langle \psi_\alpha,t|\gamma\rangle,\nn\\
&&=\frac{1}{\pi}\,e^{-(|\gamma|^2+|\eta_t|^2)}\left|\sum_{n=0}^\infty\frac{(\bar{\gamma}e^{-i\Omega_0 t}\eta_t)^n\,e^{-i\chi t n^2}}{n!}\right|^2.
\eea
\begin{figure}[ht]
\centering
\includegraphics[scale=0.18]{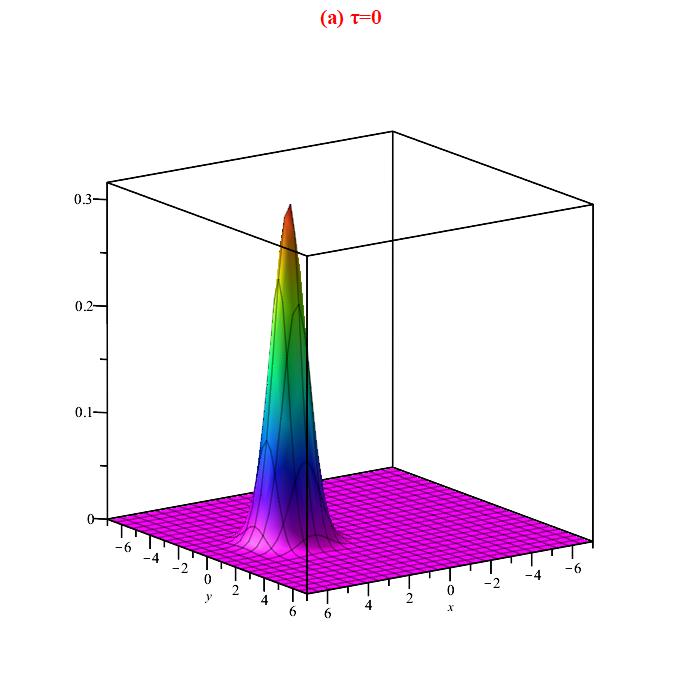}
\includegraphics[scale=0.18]{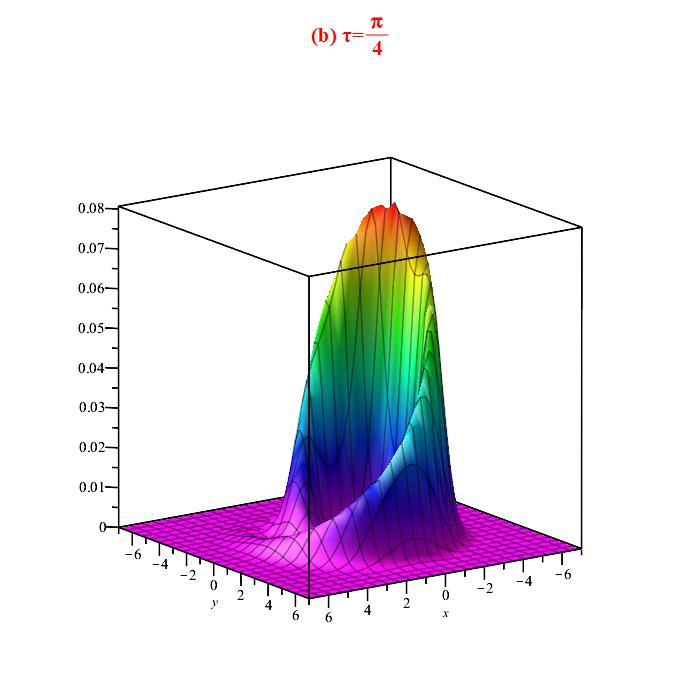}
\includegraphics[scale=0.18]{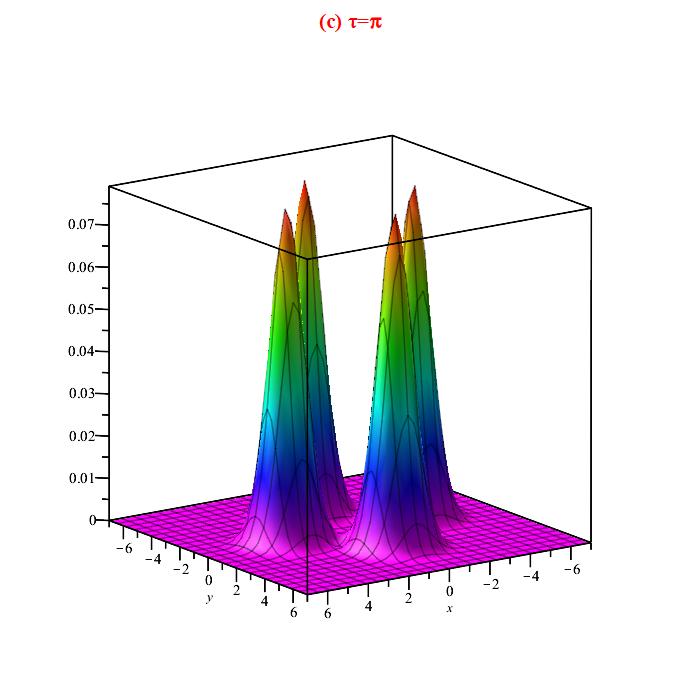}
\includegraphics[scale=0.18]{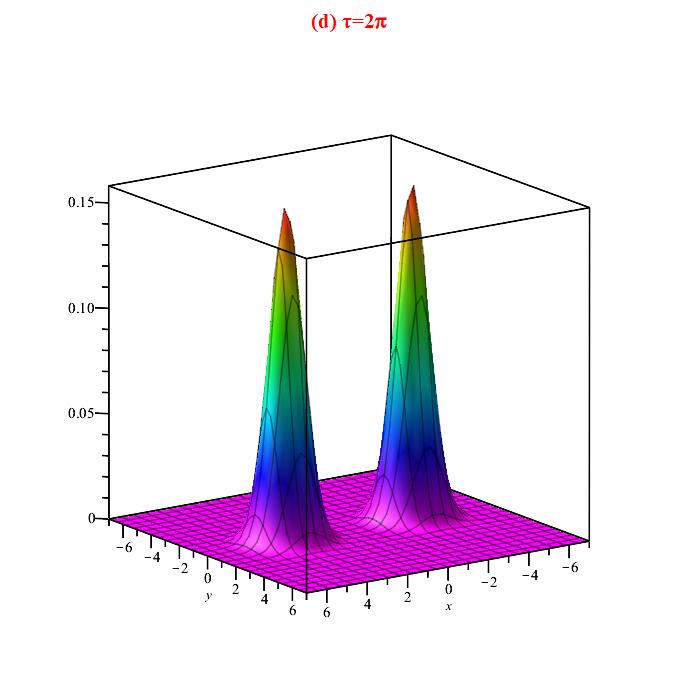}
\includegraphics[scale=0.18]{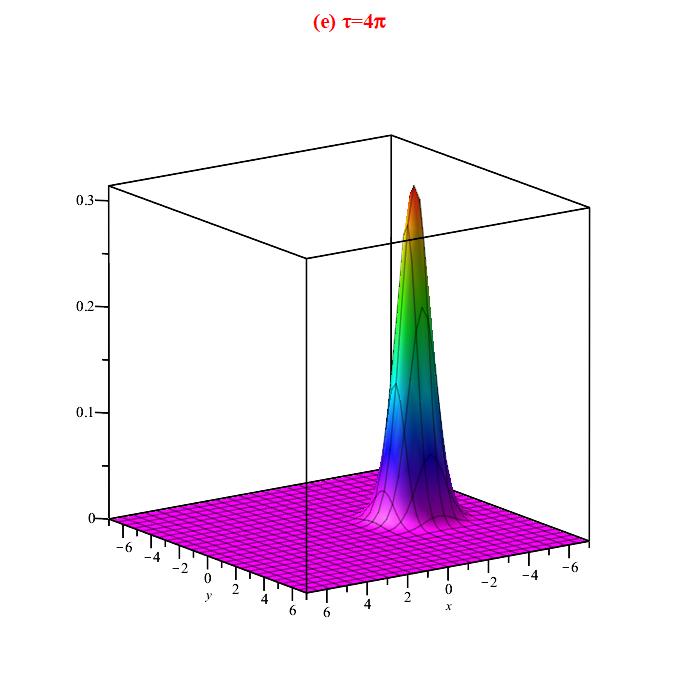}
\includegraphics[scale=0.18]{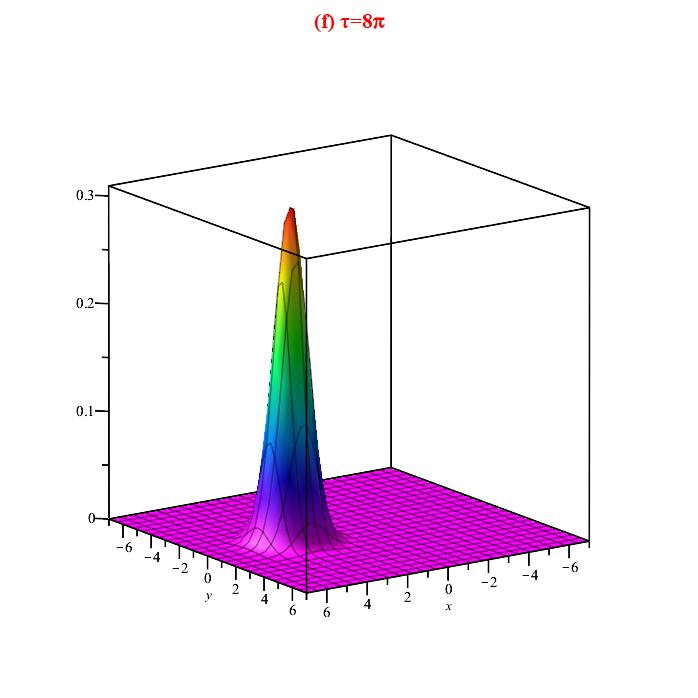}
\caption{(Color online) The Husimi distribution function depicted in $x=\mbox{Re}[\gamma]$ and  $y=\mbox{Im}[\gamma]$ at the scaled times ($\tau=\Omega_0 t=0, \pi/4, \pi, 2\pi, 4\pi, 8\pi$) for the values: $\Omega_0=1$, $\chi=0.25$, $\alpha=3$ and $e(\tau)=\cos(\tau)$. }
\label{fig2}
\end{figure}
In Fig. (\ref{fig2}), the Husimi distribution function is depicted in terms of the variables $x=\mbox{Re}[\gamma]$ and  $y=\mbox{Im}[\gamma]$ at six different times ($\tau=\Omega_0 t=0, \pi/4, \pi, 2\pi, 4\pi, 8\pi$). By using the definition $X_2(t)+\alpha=\eta_t$, at $t=0$, we have $\eta_0=\alpha$, and the Husimi distribution function represents a coherent state ($Q(\gamma,0)=e^{-|\gamma-\alpha|^2}$) with a Gaussian distribution. At the time $\tau=\pi$, Husimi distribution has four small picks, at $\tau=2\pi$, there are two picks, at $\tau=8\pi/2$, the single peak is revived but displaced in phase space, and finally, at the revival time $T_{rev}=8\pi$, the distribution is exactly revived.
%
\section*{Conclusions}
\noindent We found that the time-evolution operator of a parametric oscillator with both mass and frequency time-dependent can be obtained from the time-evolution operator of another parametric oscillator with a constant mass but time-dependent frequency followed by a time-transformation $t\rightarrow\int_0^t dt'\,1/m(t')$. We considered a driven parametric oscillator in the absence of the Kerr parameter ($\chi=0$), and by making use of a new set of ladder operators ($\hat{A}_t^\dag, \hat{A}_t$), we found the eigenvalues and eigenfunctions of the corresponding Hamiltonian $\hat{H}_f (t)$. The eigenfunctions were obtained from free eigenfunctions $\psi_n^0 (q,t)$ by shifting $q\rightarrow q-\lambda_t\sqrt{2/\Omega(t)}$, that is $\psi_n^f (q,t)=\psi_n^0 (q-\lambda_t\sqrt{2/\Omega(t)},t)$. By setting the confinement parameter $k=0$, we investigated the Hamiltonian $\hat{H}_{k=0} (t)$ perturbatively, considering the Kerr parameter $\chi$ as the perturbation parameter. Also, by linearizing the Hamiltonian, we obtained approximate solutions for the evolved ladder operators given in Eqs. (\ref{k0aadog}). We also studied the (quasi) probability distribution functions on phase space for the Hamiltonian $\hat{H}_{k=0} (t)$. The Kerr states $|\beta\ra_{\xi}$ and their relation to deformed coherent states were introduced. The photon distribution in the Kerr state $|\beta\ra_\xi$ was Poissonian and the Mandel $Q$ parameter for this state was zero since the Mandel $Q$ parameter is not sensitive to the phase of a state. The normalized variances of the position $\hat{x}=(\hat{a}+\hat{a}^\dag)/2$ and momentum $\hat{p}= (\hat{a}-\hat{a}^\dag)/2i $ were obtained and squeezing occurred only for position in the periodic short time-intervals. In the following, we found the Husimi distribution function. The Husimi distribution evolved from a single-pick state (coherent state) to a four-picks state at the scaled time $\tau=\pi$, and to a two-picks state at $\tau=2\pi$, and finally, after the revival time $T_{rev}=8\pi$, the distribution was revived.
\section*{Data availibility}
All data generated or analysed during this study are included in this published article [and its supplementary information files].
\bibliography{SciRef}
\end{document}